\def \<{\langle}
\def \>{\rangle}

\documentclass[a4paper,11pt]{article}

\usepackage{jcappub,graphicx,epsfig,natbib,color,times,bm,amsmath,multirow,hyperref, booktabs}

\usepackage{float}
\usepackage{subfig}
\usepackage{times}
\usepackage{color}
\usepackage[normalem]{ulem}
\usepackage{soul}
\usepackage[flushleft]{threeparttable}
\usepackage{ulem}
\usepackage{comment}

\newcommand{\be}{\begin{equation}}
\newcommand{\ee}{\end{equation}}
\newcommand{\bea}{\begin{eqnarray}}
\newcommand{\eea}{\end{eqnarray}}

\DeclareRobustCommand{\Sec}[1]{Sec.~\ref{#1}}

\DeclareRobustCommand{\Fig}[1]{Figure~\ref{#1}}

\DeclareRobustCommand{\Eq}[1]{Eq.~(\ref{#1})}

\DeclareRobustCommand{\r}[1]{{\rm #1}}

\newcommand{\rd}{\r{d}}

\newcommand{\mbh}{M_{\r{PBH,ini}}}
\newcommand{\fbh}{f_{\r{PBH,ini}}}
\newcommand{\alm}

\title{Investigating Primordial Black Hole Accretion through Cosmic Optical Depth}

\author[a,b]{Zi-Xuan Zhang}\emailAdd{zhangzixuan@ihep.ac.cn}
\author[c,d]{Junsong Cang}\emailAdd{cangjunsong@outlook.com}
\author[a]{Yu Gao\footnote{Corresponding authors\label{corresauth}}}\emailAdd{gaoyu@ihep.ac.cn}
\author[a]{Hong Li\textsuperscript{\ref{corresauth}}}\emailAdd{hongli@ihep.ac.cn}

\affiliation[a]{Key Laboratory of Particle Astrophysics, Institute of High Energy Physics, Chinese Academy of Sciences, 19B Yuquan Road, Beijing, China}
\affiliation[b]{School of Physical Sciences, University of Chinese Academy of Sciences, 19A Yuquan Road, Beijing, China}
\affiliation[c]{Theoretical and Scientiﬁc Data Science, Scuola Internazionale Superiore di Studi Avanzati (SISSA), Via Bonomea 265, 34136 Trieste, Italy}
\affiliation[d]{Cosmology Group, Scuola Normale Superiore,
Piazza dei Cavalieri 7, 56126 Pisa, Italy}
    
\begin{document}

\abstract{
Primordial black holes (PBH) accretion in the late Universe can lead to significant mass growth. 
A larger mass further accelerates the accretion radiation output for PBHs with initial masses greater than one solar mass, 
potentially leading to a stringent energy-dumping constraint derived from observations of the cosmic microwave background. 
The energy injected via PBH accretion is capable of ionizing and heating the intergalactic medium (IGM), 
ultimately affecting the optical depth of cosmic reionization and the 21-cm signal. 
This work investigates primordial black hole mass growth using the Bondi-Hoyle accretion model and accounts for additional ionization and heating induced by PBHs.
We derive stringent PBH abundance limits using an upper limit on optical depth set by {\it Planck} 2018 CMB measurements.
We find that accretion growth significantly strengthens late-time observational constraints for primordial black holes with initial masses ranging from several solar masses up to $10^4$ solar masses. 
The PBH fraction of the Universe's unobserved mass content can be constrained to $f_{\rm PBH, ini}\sim 10^{-2}$ to 
$10^{-7}$ in this mass range,
and when accounting for mass
evolution our constraints can be strengthened by up to one order of magnitude.
In addition, we show that PBH mass growth will lead to an observable impact on the predicted hydrogen 21-cm brightness temperature.
}
\maketitle

\section{Introduction}\label{sec:intro}

Primordial black holes (PBHs), first proposed by Carr and Hawking in the 1970s~\cite{Hawking}, can be formed as a result of the collapse of density fluctuations in the early Universe. 
Depending on the time at which fluctuation modes enter the horizon, as well as later growth~\cite{DeLuca:2020bjf}, primordial black holes may emerge in a wide range of masses~\cite{Carr:1975qj, Carr_2021}. 
One very intriguing aspect of PBHs is their potential role in the late evolution of the Universe~\cite{chapline_cosmological_1975}: as massive PBHs barely lose mass via Hawking radiation, they may survive to the present day, thereby contributing to the mass budget that influences galactic dynamics~\cite{chapline_cosmological_1975}. PBHs can also influence large-scale structure formation in the Universe, affecting matter distribution~\cite{Afshordi_2003} and many other cosmic evolution features, see Ref~\cite{Carr_2024} for a comprehensive review.

Primordial black holes have been extensively studied as candidates for dark matter (DM), and they are subject to stringent constraints based on astronomical observations.
Various methods, including PBH Hawking radiation \cite{Carr_2010, Clark:2018ghm, Acharya_2020, Cang_2022}, gravitational lensing \cite{Smyth_2020, Griest_2014, Alcock_2001, Tisserand_2007, Zumalac_rregui_2018, Niikura:2019kqi}, 
cosmic microwave background (CMB)
anisotropy \cite{Ricotti_2008, Clark:2016nst, chen_constraint_2016,poulin_cmb_2017,serpico_cosmic_2020, Facchinetti:2022kbg, Agius:2024ecw}, gravitational wave \cite{Ali_Ha_moud_2017, Abbott_2019_b, Jangra_2023, Miller_2024, Andres_Carcasona:2024wqk} and some dynamic methods \cite{Capela_2013, Monroy_Rodr_guez_2014, Graham_2015} have been adopted to test PBHs as dark matter in different mass ranges. Projects such as MACHO \cite{Alcock_2001} and EROS \cite{Tisserand_2007} have utilized gravitational microlensing to search for PBHs with masses between $10^{-7}$ and $10$ solar masses. 
Recently, studies using data from the {\it Planck} satellite have constrained the abundance of PBHs in the mass range of 10 to $10^4$ solar masses \cite{serpico_cosmic_2020}. Even at a sub-unity fraction of the invisible mass budget, PBHs are still of strong interest due to other roles in astrophysics and cosmology,
e.g.
PBHs are predicted to heat the galaxy cores~\cite{Bullock_2017}, seed supermassive black holes~\cite{Bean_2002} and galaxies~\cite{Yuan:2023bvh}, and generate large-scale structure through Poisson fluctuations~\cite{Afshordi_2003}, etc.

The detection of gravitational waves from binary black hole mergers by LIGO and Virgo has revitalized interest in massive PBHs~\cite{Bird:2016dcv, Clesse:2016vqa, Sasaki:2016jop, Abbott_2016, Abbott_2016_2, Abbott_2016_3, Garc_a_Bellido_2017, Franciolini_2022}. Both binary PBHs, along with astrophysical populations, may contribute to explaining the observed gravitational wave events. Substantial efforts are underway to investigate the fundamental characteristics of PBHs, including their spin~\cite{Bhagwat_2021}, torque~\cite{Chen_2018}, and mass distributions~\cite{Chen_2022}.

The observed black hole sizes in LIGO binary merger events suggest that they may have experienced efficient accretion growth during cosmic history~\cite{De_Luca_2021, DeLuca:2020bjf}. 
Ref.~\cite{De_Luca_2020} showed that mass accretion can significantly weaken PBH abundance constraints today.
Additionally, primordial black holes may have served as seeds for early galaxy formation through accretion, making it essential to consider this process when examining massive PBHs in the later stages of the universe.
The rate and nature of accretion can provide critical insights into the growth of PBHs and their potential detectability. Ref.~\cite{De_Luca_2020} demonstrated that the mass evolution of PBHs can be significant in the mass range [1, $10^4]\ M_\odot$. However, many studies of PBHs within this mass window have yet to incorporate the effects of mass evolution. In this paper, we will focus on the mass evolution of PBHs induced by accretion, particularly during the crucial periods before and after the onset of cosmic reionization.

Prior to the epoch of reionization (EoR), X-ray emissions from the heated gas during PBH accretion are capable of ionizing and heating up the gases in the intergalactic medium, causing an increase in the ionization fraction $x_\mathrm{e}$ and gas temperature~\cite{Slatyer_2013}. A higher medium ionization fraction contributes to random scattering of the CMB photons, leading to the smearing of the CMB's anisotropy correlation and increased optical depth~\cite{Padmanabhan:2005es}. Both ionization and temperature rise in the intergalactic medium will reduce the strength of 21cm signal during the pre-EoR window~\cite{1952AJ.....57R..31W, 4065250}. When PBHs are allowed to grow, their accretion radiation can become much more efficient towards the time of reionization and a stronger impact can be expected on the CMB and the potential 21cm signal.        

Here we use the optical depth constraints derived from the {\it Planck} CMB data to constrain PBH ionization effects and derive upper limits on PBH abundance.
The rest of this paper is organized as follows: Section~\ref{sec:abh} discusses the accretion model for PBHs, Section~\ref{sec:effects} reviews the effects of PBHs on the intergalactic medium. We present our results derived from corrections on the optical depth in Section~\ref{sec:result} and finally we summarize in Section~\ref{sec:conclusions}.
Throughout this work we use $\Lambda \r{CDM}$ cosmological parameters set by {\it Planck} 2018 results~\cite{2020}:
$H_0 = 67.66\ \r{kms^{-1}Mpc^{-1}}$,
$\Omega_{\Lambda} = 0.6903$,
$\Omega_\r{M} = 0.3096$,
$\Omega_\r{C} = 0.2607$,
$\Omega_\r{B} = 0.0489$,
$\ln(10^{10}A_\r{s}) = 3.047$,
$n_\r{s} = 0.967$.

\section{Accreting Black Holes}\label{sec:abh}
\subsection{The accretion rate of PBHs}\label{sec:abh_mdot}

After formation,
PBHs may accrete surrounding baryonic matter, and this can potentially lead to substantial growth of PBH mass.
Ignoring the angular momentum of the accreted gas,
gas accretion rate for a point mass $M$ can be computed with the Bondi-Hoyle rate $\dot{M}_{\mathrm{B}}$~\cite{10.1093/mnras/104.5.273},
a PBH with mass $M_\mathrm{PBH}$ moving at a speed $v_\mathrm{rel}$ in a homogeneous gas with number density $n_\mathrm{gas}$ and sound speed $c_\mathrm{s}$ can accrete at the Bondi-Hoyle rate $\dot{M}_{\mathrm{B}}$~\cite{10.1093/mnras/104.5.273},
\begin{align}
    \dot{M}_{\mathrm{B}} (M) =  4 \pi \lambda M n_{\mathrm{gas}} v_{\mathrm{eff}} r_\mathrm{B}^2,
\label{eq:Bondi_rate}
\end{align}
where $n_\mathrm{gas}$ is the gas number density,
the Bondi-Hoyle radius $r_\r{B}$ is given by,
\be
r_\mathrm{B}(M) = GM/v_\mathrm{eff}^2,
\ee
and 
$v_\mathrm{eff} \equiv \big<(c_\mathrm{s}^2 + v_\mathrm{rel}^2)^{-3}\big>^{-1/6}$
is the PBH effective velocity,
determined by the gas sound speed $c_\mathrm{s}$ and the relative velocity $v_\mathrm{rel}$ between PBHs and baryons. In the linear regime, the square root of the variance of the $v_\mathrm{rel}$ can be expressed as~\cite{Yang_2021},
\begin{align}
    \big<v_{\mathrm{rel}}^2\big>^{1/2} \approx \mathrm{min}[1, z/10^3]
    \times \mathrm{30\ km\ s^{-1}},
\label{eq:vrel}
\end{align}
and $c_\r{s}$ can be written as~\cite{poulin_cmb_2017},
\begin{align}
    c_\r{s} = \sqrt{\frac{\gamma(1+x_\r{e})T_\r{K}}{m_\r{p}}},
\label{eq:sound_speed}
\end{align}
where $\gamma=5/3$, $m_\r{p}$ is the proton mass,
$x_\r{e} \equiv n_\r{e}/n_\r{H}$ is the ionisation fraction,
$n_\r{e}$ and $n_\r{H}$ denote number densities of free electron and hydrogen nuclei,
respectively,
$T_\r{K}$ is the kinetic temperature of gas.

Finally in \Eq{eq:Bondi_rate},
$\lambda$ is a dimensionless accretion parameter that accounts for the effects of Hubble expansion and Compton drag.
A good fit to $\lambda$ is given by Ricotti \cite{Ricotti_2007},
\begin{align}
    \lambda = \mathrm{exp}\bigg(\frac{9/2}{\hat{\beta}^{0.75}+3}\bigg)x_{\mathrm{cr}}^2,
\label{eq:lambda_naked}
\end{align}
where the dimensionless sonic radius $x_\mathrm{cr}$ can be written as,
\begin{align}
    x_{\mathrm{cr}} = [-1 + (1+\hat{\beta})^{1/2}]/\hat{\beta}.
\label{eq:sonic_radius}
\end{align}
here $\hat{\beta}$ is the effective gas viscosity,
\begin{equation}
\begin{aligned}
    \hat{\beta}(M) &= \bigg(\frac{M}{10^4M_\odot}\bigg)
    \bigg(\frac{1+z}{1000}\bigg)^{3/2}
    \bigg(\frac{c_\mathrm{s}}{5.74\ \mathrm{km\ s^{-1}}}\bigg)^{-3}\\
    &\times
    \bigg[0.257 + 1.45\bigg(\frac{x_\mathrm{e}}{0.01}\bigg)\bigg(\frac{z+1}{1000}\bigg)^{5/2}\bigg].
\end{aligned}
\label{eq:eff_viscosity}
\end{equation}

Note that at lower redshifts,
DM particles can be gravitationally attracted towards PBHs where they collapse into DM halos,
and this can impact PBH accretion rate.
The density profile of DM halos formed around PBHs follows a power law of the form $\rho\sim r^{-\alpha}$,
where $\alpha \sim 2.25$ \cite{Ricotti_2007} and $r$ is the distance towards the halo center,
and the DM halo mass $M_\r{h}$ is related to PBH mass $M_\r{PBH}$ and redshift $z$ by,
\begin{align}
    M_\mathrm{h} = 3M_\mathrm{PBH}\bigg(\frac{1000}{1+z}\bigg),
\label{eq:halo_mass}
\end{align}
and the radius of the halo is,
\begin{equation}
    r_\mathrm{h} 
    =0.019\bigg(\frac{M_\mathrm{h}}{M_\odot}\bigg)^{\frac{1}{3}}\bigg(\frac{1+z}{1000}\bigg)^{-1}\mathrm{pc}.
\label{eq:halo_radius}
\end{equation}

When the ratio parameter,
\begin{align}
    \chi \equiv \frac{r_\mathrm{B}}{r_\mathrm{h}},
\label{eq:Chi}
\end{align}
satisfies $\chi \ge 2$,
namely, the PBH Bondi-Hoyle radius is larger than twice the halo radius,
and one can treat the accretion process of the dark matter halo as a point mass with $M_\mathrm{h}$. 
Otherwise, at $\chi < 2$ when only a fraction of the halo mass participates in the Bondi-Hoyle accretion,
it is necessary to modify the relevant parameters of the point mass accretion model \cite{Ricotti_2008} following,
\begin{align}
    \hat{\beta}^{(h)} \equiv \chi^{\frac{p}{1-p}} \hat{\beta}(M_\r{h}),\ \ 
    \lambda^{(h)} \equiv \Upsilon^{\frac{p}{1-p}} \lambda,\ \ 
    x_\r{cr}^{(h)}=\left(\frac{\chi}{2}\right)^{p/(1-p)}x_\r{cr}(\hat{\beta}^{(h)}),
\label{eq:modified_para}
\end{align}
where $p\sim 0.75$ and,

\begin{align}
    \Upsilon = \bigg(1+\hat{\beta}^{(h)}\bigg)^{\frac{1}{10}}\mathrm{exp}(2-\chi)\bigg(\frac{\chi}{2}\bigg).
\label{eq:modified_upsilon}
\end{align}

Note that the accretion of dark matter to PBHs is negligible~\cite{Ricotti_2008} and thus not accounted for in this work.
The velocity dispersion of dark matter is typically much higher than the escape velocity required for dark matter to be gravitationally bound to the PBH,
this dynamical imbalance implies that the random thermal motions of DM particles dominate over the PBH's gravitational pull,
rendering direct accretion inefficient.
Furthermore,
the density profile of the DM halo extends over scales orders of magnitude larger than the Schwarzschild radius of the PBH.
Such vast spatial separation introduces significant angular momentum barriers,
and for DM to spiral into the PBH they must first lose nearly all their angular momentum,
which is highly improbable given the lack of efficient dissipative mechanisms for DM.

\subsection{The accretion luminosity}
\label{sec:abh_lumi}

Black holes are typically found to exhibit strong X-ray emission~\cite{Gilfanov_2009} correlated with their accretion rate.
For convenience we define the dimensionless PBH accretion rate $\dot{m}$ as,
\be
\begin{aligned}
    \dot{m} &= \dot{M}_\r{PBH,acc} / \dot{M}_\r{Edd}
    \\
    & = 0.125 \lambda \frac{\Omega_\mathrm{b}}{m_\mathrm{H}}\bigg(\frac{1+z}{1000}\bigg)\bigg(\frac{M_\mathrm{PBH}}{M_\odot}\bigg)\bigg(\frac{v_\mathrm{eff}}{5.74 \mathrm{km\ s^{-1}}}\bigg)^{-3}.
\end{aligned}
\label{eq:mdot}
\ee
where $\dot{M}_\r{PBH,acc}$ is the accretion rate of PBH, 
$\dot{M}_\mathrm{Edd} \equiv L_\mathrm{Edd}/c^2$ is the Eddington accretion rate and $L_\mathrm{Edd}$ is the Eddington luminosity,
\be
L_\r{Edd} (M_\r{PBH})
=
1.3 \times 10^{38}
\left(
\frac{M_\r{PBH}}{M_\odot}
\right)
\r{erg\ s^{-1}},
\ee
The bolometric accretion luminosity $L_\r{bol}$ is related to accretion rate $\dot{m}$ by,
\be
L_\r{bol}
=
\eta
\dot{m}
L_\r{Edd},
\label{Eq_Lbol_PBH}
\ee
where the radiative efficiency $\eta$ depends on both accretion rate and the geometry of accretion flow~\cite{Zhang:2023rnp}.
Depending on the accretion rate,
accrected gas forms either an accretion disk ($\dot{m} \ge 1$) or an advection-dominated accretion flow (ADAF, $\dot{m} < 1$)~\cite{Narayan:2021qfw,Yuan:2014gma,Hektor:2018qqw}.
$\eta$ is expected to be proportional to $\dot{m}$ at low accretion rate,
whereas for $\dot{m}\ge0.1$ it approaches the constant value of 0.1 set by disk accretion geometry.
Following~\cite{Zhang:2023rnp,Narayan:2008bv},
we model $\eta$ as,
\be
\eta
=
\min(\dot{m},0.1).
\label{Eq_Eta}
\ee
We adopt the Eddington luminosity as the upper limit for PBH bolometric luminosity,
\be
L_\r{bol}
\le
L_\r{Edd}
\label{Eq_Eddington_Limit}
\ee
which limits $\dot{m} \le 10$.

Motivated by fits to theoretical calculations of ADAF accretion~\cite{Merloni:2003aq} and X-ray luminosity observations from the XMM-COSMOS survey~\cite{Lusso:2012yv},
we follow~\cite{Zhang:2023rnp} to model PBH X-ray emission as,

\begin{equation}
    \mathrm{log}_{10}(L_\mathrm{X}/L_\mathrm{Edd})=\left\{
    \begin{aligned}
    &2.3 ~\mathrm{log}_{10}\dot{m} + 1.1 \ \ \ \  &\dot{m} \leq 0.02\\
    &0.25 ~\mathrm{log}_{10}\dot{m} -2.4&\dot{m} >0.02
   \end{aligned}
   \right.
   .
\end{equation}
where $L_\r{X}$ is the bolometric X-ray luminosity,
\begin{align}
    L_\mathrm{X} = \int_{2~\mathrm{keV}}^{10~\mathrm{keV}}\mathcal{L}_\mathrm{X}(E){\bf d}E.
\label{eq:lumiosity}
\end{align}
and $\mathcal{L}_\r{X} \equiv \rd L_\r{X} /\rd E$ is the specific X-ray luminosity,
which follows an energy spectrum of form~\cite{Zhang:2023rnp},
\begin{align}
    \mathcal{L}_\mathrm{X}(E) 
    \propto \bigg(\frac{E}{2~\mathrm{keV}}\bigg)^{-0.5}
    \mathrm{exp}\bigg(-\frac{E}{100~\mathrm{keV}}\bigg),
\label{Eq_LX_Specific}
\end{align}

\subsection{Mass evolution and boosts to radiation output and accretion rate}

Accretion can lead to significant growth in PBH mass.
The energy carried away from PBH by accretion luminosity outflow (see \Eq{Eq_Lbol_PBH}) does not contribute to mass evolution,
therefore ignoring the angular momentum of gas,
the PBH mass growth rate $\dot{M}_\r{PBH}$ is related to both accretion rate (gas inflow) $\dot{m}$ and radiative efficiency $\eta$ by,
\be
\dot{M}_\r{PBH} = (1 - \eta) \dot{m}\dot{M}_\r{Edd}.
\label{Eq_dMPBH_dt}
\ee
For sufficiently large accretion rate and PBH abundance,
PBH accretion may consume a substantial fraction of baryonic matter and this can in turn suppress the accretion rate.
For the regime of interest in this work,
we have checked that the fractional loss in baryonic gas is negligible (at up to $10^{-2}$ level),
thus we ignore this effect in our subsequent analysis.

In the most extreme case in which $\dot{m}$ consistently takes the upper limit of 10 set by Eddington limit in \Eq{Eq_Eddington_Limit},
\Eq{Eq_dMPBH_dt} can be solved analytically and it can be shown that fractional mass evolution $M_\r{PBH}/\mbh$ is dependent only on the age $t$ of the Universe
\footnote{
We caution our readers that \Eq{Eq_MassEvo_Edd_Limit} is only valid for small PBH mass density $\rho_\r{PBH}$.
As we will discuss in detail at next section,
if $\rho_\r{PBH}$ were to be large enough,
X-ray from PBH accretion can increase IGM temperature and ionisation,
and this may in turn suppress accretion rate (see e.g. \Eq{eq:sound_speed}) $\dot{m}$ and then \Eq{Eq_MassEvo_Edd_Limit} will no longer be valid.
},
\be
\frac{M_\r{PBH}}
{\mbh}
=
\exp\left(
\frac{285.2t}{t_0}
\right),
\ \r{Eddington\ limit}
\label{Eq_MassEvo_Edd_Limit}
\ee
where $t \in [0, t_0]$,
corresponding to redshift range of $[0, \infty]$.
$\mbh$ is the initial PBH mass,
$t_0 = 13.8$ Gyr is the current age of the Universe.
When computing mass evolution for realistic accretion scenarios detailed in \Sec{sec:abh_mdot},
we integrate \Eq{Eq_dMPBH_dt} from redshift $z=10^3$.
We clarify that here the {\it initial} values of both PBH mass and abundance are evaluated at $z\ge10^3$,
although we note that they may also be extended to $z \sim 350$ because \Eq{Eq_MassEvo_Edd_Limit} shows that even for the most extreme case,
PBH mass will only evolve by about 5\% at redshift 350.

\begin{figure}
\begin{center}
{\includegraphics[width=\textwidth]{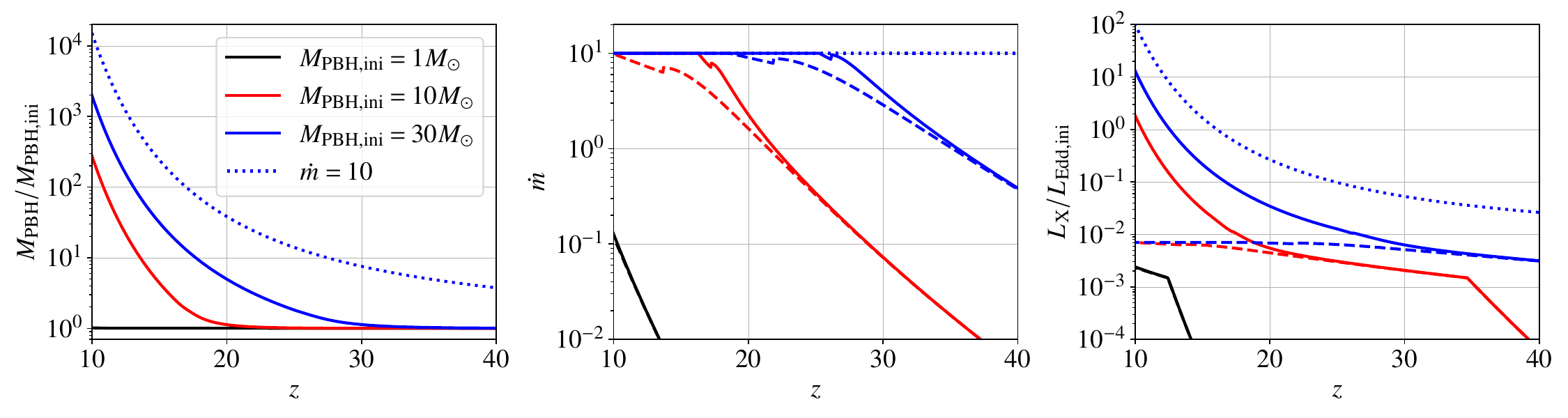}}        
\end{center}
\caption{
Evolution of PBH mass (left),
dimensionless accretion rate ($\dot{m}$, middle) and X-ray luminosity (right).
The black, red and blue solid curves represent initial PBH mass $M_\r{PBH, ini}$ of 1 $M_\odot$, 10 $M_\odot$ and 30 $M_\odot$, respectively,
the dashed lines show results when the mass evolution is not accounted for.
We normalize the evolution of PBH masses $M_\r{PBH}$ by their respective initial values $M_\r{PBH, ini}$,
and X-ray luminosity $L_\r{X}$ is normalized to initial Eddington luminosity $L_\r{Edd, ini} \equiv L_\r{Edd}(M_\r{PBH, ini})$.
The blue dotted lines show the extreme case in which PBH with $M_\r{PBH, ini} = 30 M_\odot$ accretes consistently at the upper limit of $\dot{m} = 10$.
}
\label{fig:evo}
\end{figure}

\Fig{fig:evo} showcases the impacts of accretion on PBH mass and X-ray output.
From left to right,
we show the evolutions of PBH mass, accretion rate and X-ray luminosity, respectively.
We consider three different initial PBH masses $M_\mathrm{PBH, ini}$ of 1, 10 and 30 $M_\odot$.
For convenience,
in \Fig{fig:evo} we assume that the initial mass density of PBH is a fraction of $10^{-8}$ of that of DM,
such that the relevant IGM temperature and ionization remains the same as the $\Lambda \r{CDM}$ case with no PBH.
Due to observational uncertainties in IGM thermal and ionization,
we only show results for $z \ge 10$.

From the middle panel of \Fig{fig:evo} it can be seen that for the small initial mass of 1 $M_\odot$,
the accretion $\dot{m}$ remains between $[10^{-5},10^{-1}]$,
which is far below the threshold for appreciable mass increase,
thus in the left panel we see that the relevant mass remains unchanged.
We find that heavier PBHs generally exhibit higher accretion rate $\dot{m}$ until it saturates at the Eddington limit$\dot{m}\le 10$,
thus as the initial PBH mass increases,
the mass growth becomes more significant.
With initial mass of 30 $M_\odot$,
PBH mass can increase by a factor of $2 \times 10^3$ by redshift 10.
For the most extreme case in which PBHs consistently accrete at the upper limit of $\dot{m} = 10$,
as shown in the blue dotted lines,
PBH mass can grow by a factor of $1.7 \times 10^4$ at $z=10$
(this number can also be derived from \Eq{Eq_MassEvo_Edd_Limit} using $t_{z=10} = 0.0342t_0$).

To highlight the impact of mass evolution,
in the middle and right panels of \Fig{fig:evo},
we also show results without accounting for mass evolution such that $M_\r{PBH} = M_\r{PBH,ini}$ throughout (dotted lines).
It can be seen from the red and blue curves that for the same initial mass,
as PBH gains more mass,
the relevant accretion rate also increases compared with the case of no evolution,
i.e. accretion increases not only PBH mass but also accretion rate itself,
and this can have a huge impact on accretion X-ray outputs.

In the right panel of \Fig{fig:evo},
we show accretion X-ray luminosity $L_\r{X}$ normalized by the initial Eddington luminosity $L_\r{Edd, ini} \equiv L_\r{Edd}(M_\r{PBH, ini})$.
Similar to the results for mass evolution and accretion rate,
we find that heavier PBHs exhibit higher luminosity.
More importantly,
once PBH starts to gain mass,
the relevant X-ray luminosity shows sharp increase relative to the case of no mass evolution.
For $M_\r{PBH, ini} = 30 M_\odot$,
X-ray radiation output can be boosted by a factor of 2000,
and for the extreme $\dot{m}=10$ case (blue dotted line),
the increase can exceed 4 orders of magnitude.

\section{Impact on the IGM}
\label{sec:effects}

Radiation from accreting PBHs propagates through the Universe and gradually transfers its energy into the intergalactic medium (IGM) through occasional collisions with the intergalactic gas. 
Such energy sources will modify the evolution of the IGM temperature $T_\mathrm{K}$ and the ionization fraction $x_\mathrm{e}$ during the cosmic dark age.
This section details how energy is injected from accreting PBH and subsequently absorbed by IGM.

In the scope of this paper,
the energy released by accreting PBHs is absorbed by the IGM primarily through three deposition channels: IGM heating (Heat), hydrogen ionization (HIon), and excitation (Ly$\alpha$). In the presence of these additional energy deposition processes, the modifications to the standard evolution equations for $T_\mathrm{K}$ and $x_\mathrm{e}$ take the following form~\cite{PhysRevD.95.023010, PhysRevD.94.063507, PhysRevD.102.103005}

\begin{align}
    \frac{{\bf d}T_\mathrm{K}}{{\bf d}t} = \bigg[\frac{{\bf d}T_\mathrm{K}}{{\bf d}t}\bigg]_0
    + \frac{2}{3n_\mathrm{H}(1+f_\mathrm{He}+x_\mathrm{e})}\bigg[\frac{{\bf d}E}{{\bf d}V{\bf d}t}\bigg]_\mathrm{dep,Heat},
\label{eq:history_xe}
\end{align}
\begin{align}
    \frac{{\bf d}x_\mathrm{e}}{{\bf d}t} = \bigg[\frac{{\bf d}x_\mathrm{e}}{{\bf d}t}\bigg]_0
    + \frac{1}{3n_\mathrm{H}(z)E_\mathrm{i}}\bigg[\frac{{\bf d}E}{{\bf d}V{\bf d}t}\bigg]_\mathrm{dep,HIon}
    +\frac{1-C}{n_\mathrm{H}(z)E_\mathrm{\alpha}}\bigg[\frac{{\bf d}E}{{\bf d}V{\bf d}t}\bigg]_\mathrm{dep,Ly\alpha},
\label{eq:history_Tk}
\end{align}
where $[\rd E/\rd V \rd t]_\r{dep, c}$ represents the PBH X-ray energy output deposited to IGM per unit time and volume,
$\r{c}\in[\r{Heat, HIon, Ly\alpha}]$ denotes different deposition channels.
$n_\r{H}$ is the number density of hydrogen nuclei.
The terms with subscript $0$ represent the evolution equations in standard $\Lambda$CDM cosmology~\cite{Seager:1999bc, Ali-Haimoud:2010hou},
$E_{\r{i}} = 13.6 \r{eV}$ is the ionisation energy of neutral hydrogen, while 
$E_{\alpha} = 10.2$ eV is the excitation energy to hydrogen's first excited state. The Peebles factor $C$ describes the probability of an excited hydrogen atom transitioning back to its ground state, and 
$f_{\r{He}}$ denotes the number fraction of helium nuclei.

To calculate the energy deposition of radiated X-ray photons across different channels, one can track the energy loss processes of the photons, which can be described by a transfer function $\mathcal{T}_{\r{c}}(z, E,z')$~\cite{Slatyer_2016}. 
This function represents the fraction of energy $E$ deposited into channel $c$ during unit $\r{ln}(1+z)^{-1}$ interval near redshift $z$. 
It can be shown analytically~\cite{Cang_2022} that the energy deposition $[\rd E/\rd V \rd t]_\r{dep, c}$ is given by,

\begin{align}
    \bigg[\frac{{\bf d} E}{{\bf d}V{\bf d}t}\bigg]_\mathrm{dep,c} 
    =(1+z)^3H(z)\int \frac{{\bf d}z'}{(1+z')^4H(z)}\int {\bf d}E\mathcal{T}_\mathrm{c}(z,E,z')\epsilon_\r{X}(E,z'),
\label{eq:mass_distribution}
\end{align}
where $\epsilon_\r{X}$ is the specific intensity for PBH energy injection,
defined as energy injected per unit time, volume and energy interval.

Here we consider the monochromatic PBH mass distribution which assumes that all PBHs are born with the same mass $M_\r{PBH, ini}$,
in this case it can be shown that the emissivity $\epsilon_\r{X}$ takes the form,
\be
\epsilon_\r{X}
=
\mathcal{L}_\r{X}
n_\r{PBH},
\label{Eq_dNdEdVdt}
\ee
where $\mathcal{L}_\r{X}$ is PBH specific X-ray luminosity given in \Eq{Eq_LX_Specific},
$n_{\mathrm{PBH}}$ is the PBH number density, 
\begin{align}
    n_{\mathrm{PBH}}(z) = f_{\mathrm{PBH, ini}}\frac{\Omega_\mathrm{c}\rho_\mathrm{cr}(1+z)^3}{M_\mathrm{PBH, ini}},
\label{eq:_spectrum}
\end{align}
where $\Omega_\r{c}$ is the fractional density parameter of DM,
$\rho_\r{cr}$ is the critical density,
$f_\r{PBH, ini}$ is the initial ratio between PBH density $\rho_\r{PBH}$ and DM density $\rho_\r{DM}$,
\be
f_\r{PBH, ini}
\equiv
\frac{\rho_\r{PBH, ini}}
{\rho_\r{DM}},
\label{Eq_fPBH}
\ee
here $\rho_\r{PBH, ini}$ is the initial PBH mass density.
Note that while \Eq{eq:_spectrum} shows that the {\it comoving} number density of PBHs remains constant,
mass evolution due to accretion can increase both PBH comoving mass density and PBH fraction in DM.

We modified the {\tt HyRec}~\cite{Ali-Haimoud:2010hou} codes based on Eq. (\ref{eq:history_xe}) and Eq. (\ref{eq:history_Tk}) to calculate the ionization and thermal history in presence of PBH accretion.
Note that energy deposition process is dependent on IGM thermal and ionization states~\cite{Slatyer_2016}.
For example,
when IGM is already ionized one expects that most of injected energy will contribute to heating rather than contributing to ionization (see e.g.~\cite{shull1985x,Galli:2013dna}).
The transfer function $\mathcal{T}_\r{c} (z, E, z')$ we used here in \Eq{eq:mass_distribution} to track energy deposition was computed in~\cite{Slatyer_2016} assuming a $\Lambda \r{CDM}$ recombination history,
which serves as a good approximation for high redshifts.
Due to observational uncertainties about IGM at lower redshifts and subsequent energy deposition process,
in this work we truncate PBH energy injection at $z=11$,
which is when the transfer function $\mathcal{T}$ in \cite{Slatyer_2016} was also truncated.

\begin{figure}
\begin{center}
{\includegraphics[width=\textwidth]{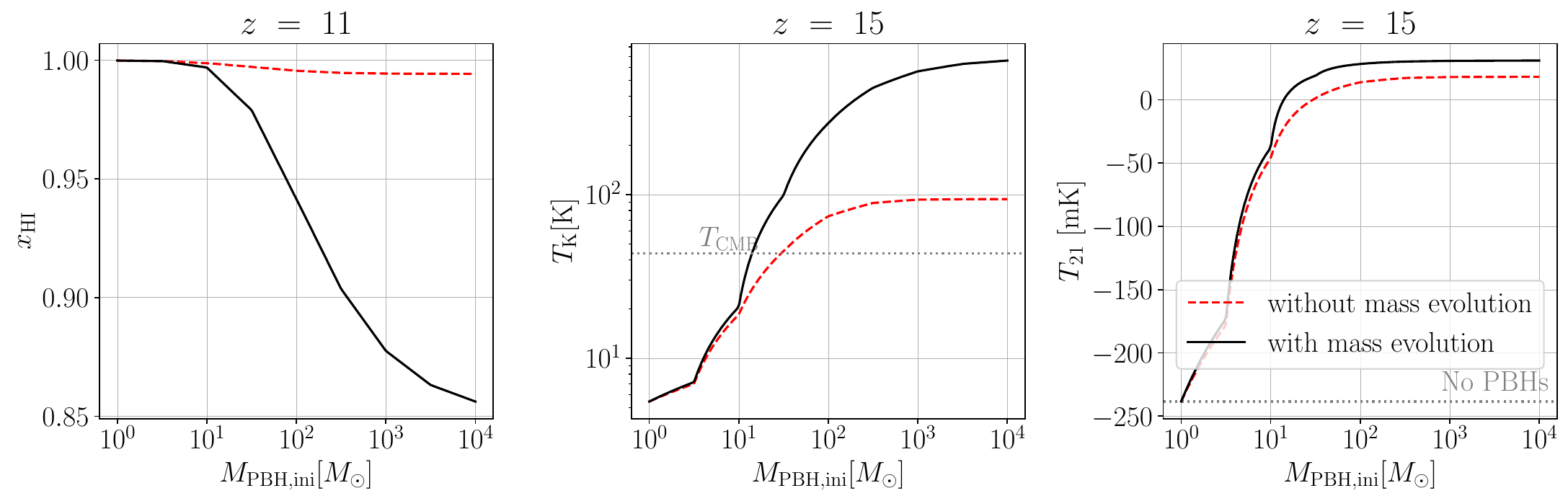}}        
\end{center}
\caption{
Neutral hydrogen fraction (left), 
gas temperature (middle), 
and 21cm brightness temperature (right)  for different initial PBH masses.
$x_{\rm HI}$ is plotted for $z=11$,
while $T_{\rm K}$ and $T_{\rm 21}$ are shown for $z=15$.
The initial PBH abundance $f_{\rm PBH, ini}$ is set to $10^{-7}$, 
which is allowed by current experimental constraints~\cite{Carr_2024} 
The black solid and red dashed lines represent scenarios with and without mass evolution, respectively. The legend applies to all panels.
The grey dotted line in the middle panel shows CMB temperature,
and in the right panel we show lowest possible $T_{21}$ in $\Lambda \r{CDM}$ (assuming no heating from PBHs or stars).
}
\label{fig:T21}
\end{figure}

\Fig{fig:T21} showcases the effects of PBH accretion on IGM,
from left to right,
we show the IGM neutral fraction $x_\r{HI} \equiv n_\r{HI}/n_\r{H}\approx1-x_\r{e}$,
gas kinetic temperature $T_\r{K}$ and hydrogen 21cm signals for initial PBH masses between 1 and $10^4\ M_\odot$.
We show results for cases with mass evolution (black solid) and without accounting for mass evolution (red dashed).
We assume an initial PBH fraction of $f_\r{PBH, ini} = 10^{-7}$,
which is consistent with current observational constraints~\cite{Carr_2024}.
From the left panel,
it can be seen that at $z=11$,
impact of PBH accretion on neutral fraction is negligible when mass evolution is ignored.
And when mass growth is accounted for,
the boosted X-ray radiation outputs (see \Fig{fig:evo}) enhances ionization and causes up to 15\% decrease in neutral fraction.
Since accretion and energy output boost more pronounced for massive PBH,
we see that heavier PBHs have larger impact on IGM neutral fraction.

We have shown previously in \Fig{fig:evo} that PBH mass growth and boost in X-ray radiation outputs occurs mainly at lower redshifts of $z\le 30$.
Cosmic hydrogen 21-cm signal is particularly suited to explore cosmic ionisation and heating history at this window and can potentially provide valuable insights regarding PBH properties.
While the impact on cosmic optical depth primarily derives from the correction on $x_{\rm e}$ history, 
an additional hydrogen 21-cm constraint can be imposed from the extra heating on the IGM gas temperature. 
After baryons kinetically decouple from the CMB, 
neutral hydrogen's hyperfine state distribution evolves by the balance of multiple spin-flip mechanisms and it is strongly affected by the gas temperature $T_{\rm K}$, 
especially after the Wouthuysen-Field process takes effect~\cite{Pritchard:2011xb}.
With extra heating on gas, 
the 21-cm prediction will change accordingly.
The intensity of cosmological 21-cm signal is typically measured by the 21-cm brightness temperature~\cite{Pritchard:2011xb}:
\begin{align}
    T_{21} = 27x_\mathrm{HI}\left(1-\frac{T_{\rm CMB}}{T_\mathrm{s}}\right) \times \left(\frac{1+z}{10}\frac{0.15}{\Omega_\mathrm{m}h^2}\right)^{1/2}\left(\frac{\Omega_\mathrm{b}h^2
    }{0.023}\right)\mathrm{mK},
\label{eq:t21}
\end{align}
where 
$T_{\rm CMB}$ is the CMB temperature,
and $T_\mathrm{s}$ is the spin temperature.
$T_\r{s}$ is coupled to both $T_\r{CMB}$ and $T_\r{K}$ through collisional and Wouthuysen-Field coupling~\cite{Pritchard:2011xb}.
As stars begin to form at low redshifts,
they emit Lyman-Alpha photons which contribute to Wouthuysen-Field effect until $T_\r{s}$ is tightly coupled to $T_\r{K}$.
It is typically expected that Wouthuysen-Field coupling saturation occurs before X-ray from stars starts to heat up the gas
(see e.g. \cite{Pritchard:2011xb, Pritchard:2006sq,Mesinger:2010ne,Cang:2024gtp}).

Though accounting for stellar X-ray emission and heating is beyond the scope of this paper,
we briefly explore the effects of PBH on 21-cm signal assuming saturated Wouthuysen-Field coupling such that $T_\r{s}=T_\r{K}$.
In the middle and right panels of \Fig{fig:T21},
we show gas temperature and 21-cm signal, respectively,
both plotted at $z=15$.
At this redshift,
it is observationally viable for astrophysical sources to produce enough Lyman-Alpha photons needed for saturated Wouthuysen-Field coupling,
while in the meantime not over-producing X-ray photons to dominate heating history.

Qualitatively speaking,
mass growth boosts PBH radiation outputs and leads to enhanced ionization and heating.
The ionization effect lowers $x_\r{HI}$ and can suppress 21-cm signal,
however from the left panel we have seen that at $z\ge11$ the effects of PBH ionization on $x_\r{HI}$ is relatively small,
thus in \Fig{fig:T21} the impact of PBH on 21-cm signal is primarily driven by heating.
From the middle panel we see that PBH heating increases with the PBH mass, 
especially for $M_\mathrm{PBH,ini} > 10 M_\odot$,
and we find that PBH mass evolution can increase gas temperature by a factor of 7 relative to the case without accounting for mass evolution.
 It is also interesting to note that PBH radiation may even heat $T_{\rm K}$ to be above $T_{\rm CMB}$, 
 potentially flipping the sign of $T_{\rm 21}$ and turn it into an emission signal ($T_{21} > 0$).
 Upcoming measurements can test this in MWA~\cite{Tiwari:2024ivw}, MeerKAT~\cite{Carucci:2024qpm} and SKA~\cite{SKA:2018ckk}, etc. 
 The full PBH-included $T_{21}$ evolution during EoR requires a more involved simulation,
 which we will postpone to a follow-up study.

\section{Results}
\label{sec:result}

X-ray radiation from PBHs can lead to enhanced ionization $x_\r{e}$.
During reionization,
the integrated impact of PBHs on $x_\r{e}$ can be measured by the reionization optical depth,
\begin{align}
\tau = c \sigma_\r{T} \int_0^{z_\r{max}} \frac{\rd z}{(1+z)H} x_\r{e} n_\r{H}
\label{eq:optical_depth}
\end{align}
where $\sigma_\r{T} = 6.65 \times 10^{-29} \mathrm{m}^2$ is the Thomson scattering cross-section.
The integration upper limit $z_\r{max}$ marks the onset of reionization and is set to 50 in this work.
$\tau$ measures the degree of opacity of the universe to ultraviolet and optical light due to the presence of free electrons during the reionization era,
indicating how effectively radiation from the CMB is scattered.

The latest {\it Planck} 2018 CMB anisotropy data constrains $\tau = 0.0544 \pm 0.0074$ at 68\% confidence interval (C.I.),
from which one can derive the 95\% C.I. $\tau$ upper limit as,

\be
\tau \le 0.0668,
\ 95\%.\r{C.I.}
\label{Eq_tau_upper_limit}
\ee
We will use this to constrain PBH abundance.
Specifically,
since PBH increases $x_\r{e}$ and $\tau$,
we iteratively vary PBH abundance to find the maximum $f_\r{PBH,ini}$ that satisfies \Eq{Eq_tau_upper_limit},
which is then used as our $f_\r{PBH,ini}$ upper limit.

\begin{figure}
\begin{center}
{\includegraphics[width=\textwidth]{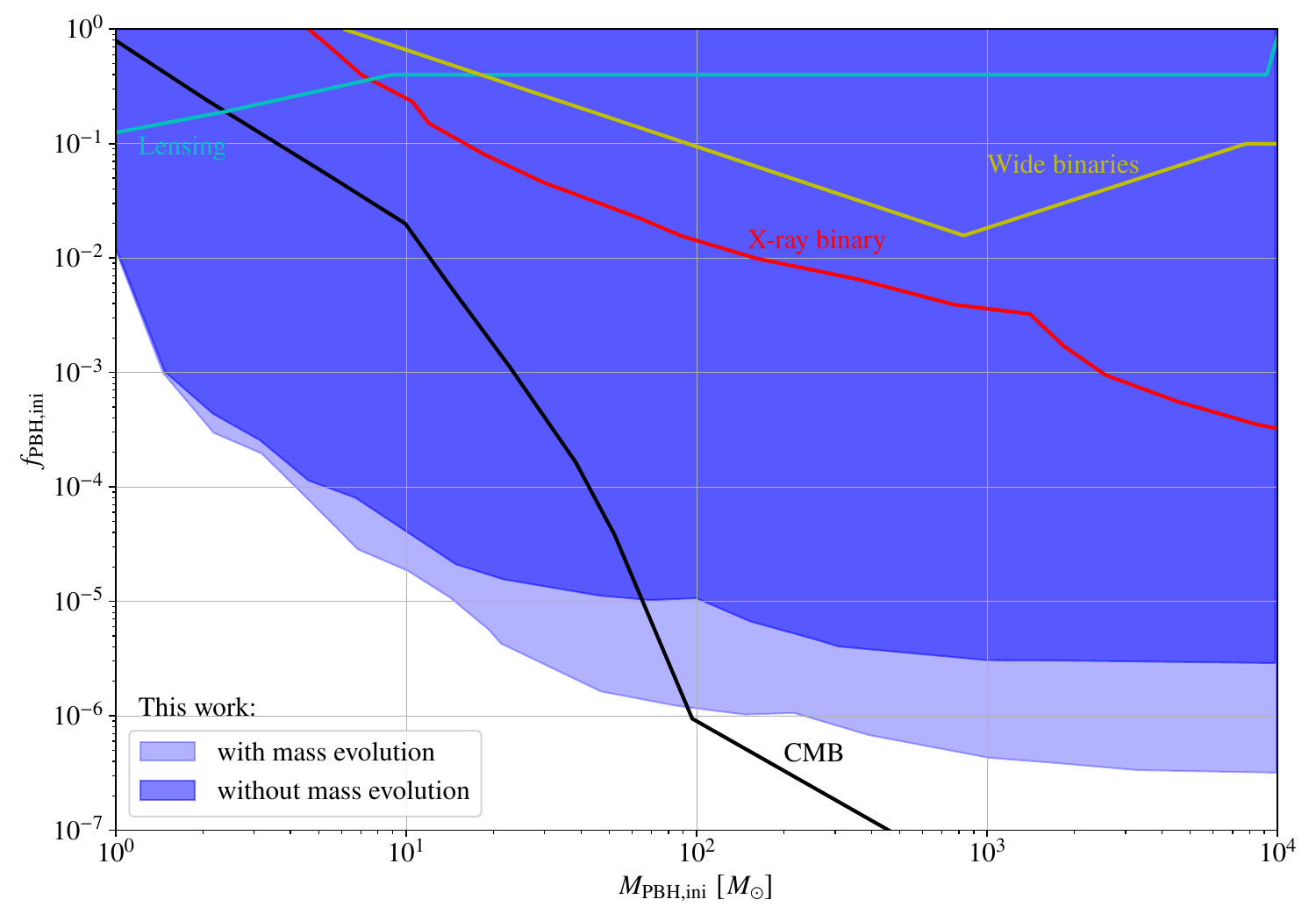}}        
\end{center}
\caption{
Upper limits on initial PBH fraction $f_\r{PBH, ini}$ derived by enforcing the {\it Planck} upper limit of $\tau < 0.0688$,
the blue dark (light) shaded region represent the case with (without) PBH mass evolution.
For comparison we also show other existing limits summarized in ~\cite{Carr:2020xqk},
specifically,
the black, red, yellow and cyan lines represent constraints set by CMB anisotropy spectrum~\cite{serpico_cosmic_2020},
X-ray binaries~\cite{inoue2017new},
wide binaries~\cite{Monroy_Rodr_guez_2014} and gravitational lensing~\cite{Tisserand_2007,niikura2019microlensing},
respectively.
}
\label{fig:results}
\end{figure}

Note that \Eq{Eq_tau_upper_limit} also constrains the contributions from astrophysical sources which are not accounted for in this work.
Furthermore as we discussed in \Sec{sec:effects},
due to observational uncertainties on reionization history which impacts PBH energy deposition,
we do not include PBH contribution when calculating $x_\r{e}$ below redshift 11,
as such our $x_\r{e}$ evolution for $z\le 11$ is driven mainly by $\Lambda \r{CDM}$ recombination terms.
Though this suggests that we underestimate $x_\r{e}$ during reionization and consequently optical depth $\tau$,
we note that when properly accounting for both these limitations,
our PBH abundance constraints will only become stronger.
As a result,
our PBH constraints derived from \Eq{Eq_tau_upper_limit} should be considered as conservative.

\Fig{fig:results} shows our upper limits on $f_\r{PBH,ini}$ for $M_\r{PBH, ini}$ within $[1, 10^4] M_\odot$ range.
Our main result is shown in the light blue shaded region,
in which we properly account for mass evolution.
From our discussions in \Sec{sec:effects},
it can be inferred that PBH contribution to ionization through deposition term $[\rd E/\rd V\rd t]_\r{dep,c}$ term is proportional to X-ray emissivity $\epsilon_\r{X}$,
and from \Eq{Eq_dNdEdVdt} and \Sec{sec:abh_lumi} one can easily show that $\epsilon_\r{X}$ follows the scaling relation,
\be
\epsilon_\r{X}
\propto
\frac{M_\r{PBH}}{M_\r{PBH,ini}}
\dot{m}^\beta,
\label{Eq_Emissivity_Scaling}
\ee
where $\beta$ equals 2.3 and 0.25 for $\dot{m} \le 0.02$ and $\dot{m} \ge 0.02$, respectively.
$\mbh$ scaling of our $\fbh$ constraints can be well explained by \Eq{Eq_Emissivity_Scaling}.
At smaller $\mbh$ when mass growth is negligible ($M_\r{PBH}/\mbh \sim 1$),
$\epsilon_\r{X}$ is dominated by $\dot{m}$ term which is found to increase with PBH mass,
thus from the black solid line we see that for $1 M_\odot \le\mbh < 10 M_\odot$,
our $\fbh$ limit tightens from $10^{-2}$ to $3\times10^{-4}$.
At higher masses,
accretion increases both PBH mass ($M_\r{PBH}/\mbh \gg1$) and accretion rate itself ($\dot{m}$),
therefore $\fbh$ limit further tightens until $\mbh \sim 10^3 M_\odot$.
At even higher masses ($\mbh \gtrsim 10^3 M_\odot$),
accretion rate $\dot{m}$ starts to approach the Eddington limit $\dot{m} = 10$.
For the most extreme case in which $\dot{m}$ consistently takes $\dot{m} = 10$,
$M_\r{PBH,ini}/\mbh$ and $\dot{m}$ terms in \Eq{Eq_Emissivity_Scaling} no longer depends on $\mbh$ (see also \Eq{Eq_MassEvo_Edd_Limit}),
thus we see that our constraints gradually flatten at $\mbh \gtrsim 10^3 M_\odot$,
reaching $\fbh \le 2\times 10^{-7}$ at $\mbh = 10^4 M_\odot$.

In the dark blue shaded region of \Fig{fig:results},
we also show constraints without accounting for mass evolution.
For this case \Eq{Eq_Emissivity_Scaling} reduces to $\epsilon \propto \dot{m}^\beta$,
since massive PBHs exhibits higher accretion rate $\dot{m}$,
our constraint tightens for higher $\mbh$ until it gradually flattens when $\dot{m}$ starts to approach the constant upper limit 10.
At $M_\odot\le \mbh < 2M_\odot$ the relevant $\fbh$ limit is identical to that of with evolution.
At higher masses,
mass evolution boosts X-ray emissivity,
therefore we see that when ignoring mass growth,
$\fbh$ constraints can be weakened by up to one order of magnitude.

Within the mass window we explored,
PBH abundance can also be constrained by other observational probes,
and we show these limits in the shaded regions for comparison,
and we note that PBH mass evolution was not accounted for in these works.
Specifically,
$\fbh$ upper limits from gravitational lensing~\cite{Tisserand_2007,niikura2019microlensing} and wide binaries range from $2\times 10^{-2}$ to 0.3 in our mass window,
and X-ray binaries~\cite{inoue2017new} observations constrain $\fbh \gtrsim 3\times 10^{-4}$ at $\mbh = 10^4M_\odot$ and weakens towards lower mass until $\mbh\sim5M_\odot$.
X-ray heating from accreting PBH can also change ionization history at much higher redshifts and leave imprints on CMB anisotropy spectrum by enhancing scattering between CMB photons and free electrons~\cite{Cang_2022,PhysRevD.95.023010,Finkbeiner:2011dx},
Ref.\cite{serpico_cosmic_2020} studied these effects using {\it Planck} CMB data and derived the previous strongest PBH limit in our mass range,
and the updated constraint we derived here is stronger than that in~\cite{serpico_cosmic_2020} by up to three orders of magnitudes in mass range $[1, 10^2]M_\odot$.
Compared with Ref.\cite{serpico_cosmic_2020} which also used CMB to study ionization effects of accreting PBHs,
here we focus on lower redshift ($z\lesssim 50$) when PBH mass evolution is non-negligible,
and we use optical depth constraints, which measures integrated PBH contribution to ionization.
In comparison,
Constraints in Ref.\cite{serpico_cosmic_2020} are derived using CMB anisotropy data,
which is only sensitive to extra energy injection at high redshifts of $z\gtrsim600$~\cite{Finkbeiner:2011dx}.
As was noted in~\cite{serpico_cosmic_2020} and can be derived from \Eq{Eq_MassEvo_Edd_Limit},
mass evolution at such high redshifts is negligible.

\section{Discussions}
\label{sec:conclusions}

Stellar mass and heavier primordial black holes may grow significantly by accreting on baryonic matter in the late Universe. 
Accretion radiation accelerates the ionization and heating of the intergalactic medium, 
leading to an observable impact on the cosmic optical depth $\tau$ and the 21-cm signal. 
Using the Bondi-Hoyle accretion model,
we study evolution of PBHs with initial mass $\mbh$ in $[1, 10^4]M_\odot$.
Massive PBHs generally exhibits higher accretion rate,
although for small solar mass PBHs the relevant mass growth is negligible,
we find that massive ($\mbh \gtrsim$) PBHs can gain substantial amount of mass through accretion.
In the most extreme cases,
accretion can increase PBH mass by up to 4 orders of magnitude by redshift 10.
Furthermore,
mass growth can also boost accretion X-ray radiation output by roughly the same orders of magnitude,
leading to enhanced ionization and heating compared to the case without accounting for evolution.

Enhanced ionization from PBHs can lead to increase in reionization optical depth $\tau$.
Using $\tau$ upper bound inferred from {\it Planck} CMB measurement,
we put limit on ionization effects of PBHs and derived stringent constraints on initial PBH abundance $\fbh$.
We constrain $\fbh \lesssim 10^{-2}$ at small mass of $\mbh = 1 M_\odot$,
and for massive PBHs our limit can tighten to $\fbh \lesssim 10^{-7}$ at $\mbh = 10^4 M_\odot$.
In mass window $[1, 10^2]M_\odot$,
our constraints are more stringent than other existing limit by up to three orders of magnitude.
We also find that enhanced radiation output due to mass growth can significantly affect PBH constraints.
When mass evolution is ignored,
our constraints can weaken by a factor of 10. 

Recently it was pointed out in Ref.~\cite{Agius:2024ecw} that the local ionization and heating can prevent the formation of DM halos surrounding the PBH,
potentially leading to reduced accretion rate.
Ref.~\cite{Facchinetti:2022kbg} further showed that the formation of ionization fronts can also suppress accretion rate as well as radiation output.
These effects can potentially weaken our PBH constraints and we reserve such analysis to future works.

Energy deposition of PBH is dependent on IGM ionization and thermal states,
due to observational uncertainties on IGM during reionization,
we do not account for PBH energy injection at $z \lesssim 11$,
and we ignored contributions from astrophysical sources for convenience.
However we note that if both these issues were to be addresses then we expect our constraints to further tighten,
thus PBH limits presented here should be considered as conservative.
PBH mass growth is expected to continue into even a lower redshift range, particularly the entire cosmic dawn and reionization epoch till around $z\sim 7$. 
Note that optical depth is an integrated quantity for CMB photon scattering and it does not fully capture the ionization history.
In the {\it Planck} 2018 results~\cite{2020},
the reionization history was modeled using a phenomenological tanh profile characterized by the reionization midpoint $z_\r{re}$.
Observational uncertainties in $z_\r{re}$ directly affects the optical depth $\tau$ constraints and consequently our PBH limits.
As shown in ~\cite{Qin:2024cna},
low-redshift observations (e.g., galaxy UV luminosity function, Lyman-Alpha forest) can help reconstruct the full reionization history and significantly tighten $z_\r{re}$ constraints compared to CMB~\cite{2020}.
Incorporating these low-redshift probes can give more robust constraints on PBHs.
Future observations are expected to yield high-precision measurements on $x_{\rm e}(z)$ history~\cite{Dai:2018nce}, 
which would provide a thorough test on the IGM environmental effect from accelerated PBH radiation.
We also note that for $\mbh>M_{\odot}$ the stronger accretion radiation also brings more pronounced suppression on the 21-cm 
absorption signal at early EoR,
and may potentially even
drive the signal to transition from absorption to emission.

\acknowledgments
The authors thank Guan-Wen Yuan, Zhihe Zhang, Yang Liu, Si-Yu Li, Yongping Li for helpful discussions. This work is supported by the NSFC under grant No. 12275278 and by the National Key R\&D Program of China No.2020YFC2201601.

\bibliography{main}

\providecommand{\href}[2]{#2}\begingroup\raggedright\begin{thebibliography}{10}

\bibitem{Hawking}
B.~J. Carr and S.~W. Hawking, \emph{{Black Holes in the Early Universe}},
  {\emph{Monthly Notices of the Royal Astronomical Society} {\bfseries 168}
  (08, 1974) 399--415}.

\bibitem{DeLuca:2020bjf}
V.~De~Luca, G.~Franciolini, P.~Pani and A.~Riotto, \emph{{The evolution of
  primordial black holes and their final observable spins}},
  \href{https://doi.org/10.1088/1475-7516/2020/04/052}{\emph{JCAP} {\bfseries
  04} (2020) 052}, [\href{https://arxiv.org/abs/2003.02778}{{\ttfamily
  2003.02778}}].

\bibitem{Carr:1975qj}
B.~J. Carr, \emph{{The Primordial black hole mass spectrum}},
  \href{https://doi.org/10.1086/153853}{\emph{Astrophys. J.} {\bfseries 201}
  (1975) 1--19}.

\bibitem{Carr_2021}
B.~Carr, K.~Kohri, Y.~Sendouda and J.~Yokoyama, \emph{Constraints on primordial
  black holes}, \href{https://doi.org/10.1088/1361-6633/ac1e31}{\emph{Reports
  on Progress in Physics} {\bfseries 84} (Nov., 2021) 116902}.

\bibitem{chapline_cosmological_1975}
G.~F. CHAPLINE, \emph{Cosmological effects of primordial black holes},
  \href{https://doi.org/10.1038/253251a0}{\emph{Nature} {\bfseries 253} (Jan.,
  1975) 251--252}.

\bibitem{Afshordi_2003}
N.~Afshordi, P.~McDonald and D.~N. Spergel, \emph{Primordial black holes as
  dark matter: The power spectrum and evaporation of early structures},
  \href{https://doi.org/10.1086/378763}{\emph{The Astrophysical Journal}
  {\bfseries 594} (Aug., 2003) L71–L74}.

\bibitem{Carr_2024}
B.~Carr, S.~Clesse, J.~García-Bellido, M.~Hawkins and F.~Kühnel,
  \emph{Observational evidence for primordial black holes: A positivist
  perspective},
  \href{https://doi.org/10.1016/j.physrep.2023.11.005}{\emph{Physics Reports}
  {\bfseries 1054} (Feb., 2024) 1–68}.

\bibitem{Carr_2010}
B.~J. Carr, K.~Kohri, Y.~Sendouda and J.~Yokoyama, \emph{New cosmological
  constraints on primordial black holes},
  \href{https://doi.org/10.1103/physrevd.81.104019}{\emph{Physical Review D}
  {\bfseries 81} (May, 2010) }.

\bibitem{Clark:2018ghm}
S.~Clark, B.~Dutta, Y.~Gao, Y.-Z. Ma and L.~E. Strigari, \emph{{21 cm limits on
  decaying dark matter and primordial black holes}},
  \href{https://doi.org/10.1103/PhysRevD.98.043006}{\emph{Phys. Rev. D}
  {\bfseries 98} (2018) 043006},
  [\href{https://arxiv.org/abs/1803.09390}{{\ttfamily 1803.09390}}].

\bibitem{Acharya_2020}
S.~K. Acharya and R.~Khatri, \emph{Cmb and bbn constraints on evaporating
  primordial black holes revisited},
  \href{https://doi.org/10.1088/1475-7516/2020/06/018}{\emph{Journal of
  Cosmology and Astroparticle Physics} {\bfseries 2020} (June, 2020)
  018–018}.

\bibitem{Cang_2022}
J.~Cang, Y.~Gao and Y.-Z. Ma, \emph{21-cm constraints on spinning primordial
  black holes},
  \href{https://doi.org/10.1088/1475-7516/2022/03/012}{\emph{Journal of
  Cosmology and Astroparticle Physics} {\bfseries 2022} (Mar., 2022) 012}.

\bibitem{Smyth_2020}
N.~Smyth, S.~Profumo, S.~English, T.~Jeltema, K.~McKinnon and P.~Guhathakurta,
  \emph{Updated constraints on asteroid-mass primordial black holes as dark
  matter}, \href{https://doi.org/10.1103/physrevd.101.063005}{\emph{Physical
  Review D} {\bfseries 101} (Mar., 2020) }.

\bibitem{Griest_2014}
K.~Griest, A.~M. Cieplak and M.~J. Lehner, \emph{Experimental limits on
  primordial black hole dark matter from the first 2 yr ofkeplerdata},
  \href{https://doi.org/10.1088/0004-637x/786/2/158}{\emph{The Astrophysical
  Journal} {\bfseries 786} (Apr., 2014) 158}.

\bibitem{Alcock_2001}
C.~Alcock, R.~A. Allsman, D.~R. Alves, T.~S. Axelrod, A.~C. Becker, D.~P.
  Bennett et~al., \emph{Macho project limits on black hole dark matter in the
  1–30 $m_\odot$ range}, \href{https://doi.org/10.1086/319636}{\emph{The
  Astrophysical Journal} {\bfseries 550} (Apr., 2001) L169–L172}.

\bibitem{Tisserand_2007}
P.~Tisserand, L.~Le~Guillou, C.~Afonso, J.~N. Albert, J.~Andersen, R.~Ansari
  et~al., \emph{Limits on the macho content of the galactic halo from the
  eros-2 survey of the magellanic clouds},
  \href{https://doi.org/10.1051/0004-6361:20066017}{\emph{Astronomy \&
  Astrophysics} {\bfseries 469} (Apr., 2007) 387–404}.

\bibitem{Zumalac_rregui_2018}
M.~Zumalacárregui and U.~Seljak, \emph{Limits on stellar-mass compact objects
  as dark matter from gravitational lensing of type ia supernovae},
  \href{https://doi.org/10.1103/physrevlett.121.141101}{\emph{Physical Review
  Letters} {\bfseries 121} (Oct., 2018) }.

\bibitem{Niikura:2019kqi}
H.~Niikura, M.~Takada, S.~Yokoyama, T.~Sumi and S.~Masaki, \emph{{Constraints
  on Earth-mass primordial black holes from OGLE 5-year microlensing events}},
  \href{https://doi.org/10.1103/PhysRevD.99.083503}{\emph{Phys. Rev. D}
  {\bfseries 99} (2019) 083503},
  [\href{https://arxiv.org/abs/1901.07120}{{\ttfamily 1901.07120}}].

\bibitem{Ricotti_2008}
M.~Ricotti, J.~P. Ostriker and K.~J. Mack, \emph{Effect of primordial black
  holes on the cosmic microwave background and cosmological parameter
  estimates}, \href{https://doi.org/10.1086/587831}{\emph{The Astrophysical
  Journal} {\bfseries 680} (jun, 2008) 829}.

\bibitem{Clark:2016nst}
S.~Clark, B.~Dutta, Y.~Gao, L.~E. Strigari and S.~Watson, \emph{{Planck
  Constraint on Relic Primordial Black Holes}},
  \href{https://doi.org/10.1103/PhysRevD.95.083006}{\emph{Phys. Rev. D}
  {\bfseries 95} (2017) 083006},
  [\href{https://arxiv.org/abs/1612.07738}{{\ttfamily 1612.07738}}].

\bibitem{chen_constraint_2016}
L.~Chen, Q.-G. Huang and K.~Wang, \emph{Constraint on the abundance of
  primordial black holes in dark matter from {Planck} data},
  \href{https://doi.org/10.1088/1475-7516/2016/12/044}{\emph{Journal of
  Cosmology and Astroparticle Physics} {\bfseries 2016} (Dec., 2016) 044--044}.

\bibitem{poulin_cmb_2017}
V.~Poulin, P.~D. Serpico, F.~Calore, S.~Clesse and K.~Kohri, \emph{{CMB} bounds
  on disk-accreting massive {Primordial} {Black} {Holes}},
  \href{https://doi.org/10.1103/PhysRevD.96.083524}{\emph{Physical Review D}
  {\bfseries 96} (Oct., 2017) 083524}.

\bibitem{serpico_cosmic_2020}
P.~D. Serpico, V.~Poulin, D.~Inman and K.~Kohri, \emph{{Cosmic microwave
  background bounds on primordial black holes including dark matter halo
  accretion}},
  \href{https://doi.org/10.1103/PhysRevResearch.2.023204}{\emph{Phys. Rev.
  Res.} {\bfseries 2} (2020) 023204},
  [\href{https://arxiv.org/abs/2002.10771}{{\ttfamily 2002.10771}}].

\bibitem{Facchinetti:2022kbg}
G.~Facchinetti, M.~Lucca and S.~Clesse, \emph{{Relaxing CMB bounds on
  primordial black holes: The role of ionization fronts}},
  \href{https://doi.org/10.1103/PhysRevD.107.043537}{\emph{Phys. Rev. D}
  {\bfseries 107} (2023) 043537},
  [\href{https://arxiv.org/abs/2212.07969}{{\ttfamily 2212.07969}}].

\bibitem{Agius:2024ecw}
D.~Agius, R.~Essig, D.~Gaggero, F.~Scarcella, G.~Suczewski and M.~Valli,
  \emph{{Feedback in the dark: a critical examination of CMB bounds on
  primordial black holes}},
  \href{https://doi.org/10.1088/1475-7516/2024/07/003}{\emph{JCAP} {\bfseries
  07} (2024) 003}, [\href{https://arxiv.org/abs/2403.18895}{{\ttfamily
  2403.18895}}].

\bibitem{Ali_Ha_moud_2017}
Y.~Ali-Haïmoud, E.~D. Kovetz and M.~Kamionkowski, \emph{Merger rate of
  primordial black-hole binaries},
  \href{https://doi.org/10.1103/physrevd.96.123523}{\emph{Physical Review D}
  {\bfseries 96} (Dec., 2017) }.

\bibitem{Abbott_2019_b}
B.~Abbott, R.~Abbott, T.~Abbott, S.~Abraham, F.~Acernese, K.~Ackley et~al.,
  \emph{Search for subsolar mass ultracompact binaries in advanced ligo’s
  second observing run},
  \href{https://doi.org/10.1103/physrevlett.123.161102}{\emph{Physical Review
  Letters} {\bfseries 123} (Oct., 2019) }.

\bibitem{Jangra_2023}
P.~Jangra, B.~J. Kavanagh and J.~Diego, \emph{Impact of dark matter spikes on
  the merger rates of primordial black holes},
  \href{https://doi.org/10.1088/1475-7516/2023/11/069}{\emph{Journal of
  Cosmology and Astroparticle Physics} {\bfseries 2023} (Nov., 2023) 069}.

\bibitem{Miller_2024}
A.~L. Miller, N.~Aggarwal, S.~Clesse, F.~De~Lillo, S.~Sachdev, P.~Astone
  et~al., \emph{Gravitational wave constraints on planetary-mass primordial
  black holes using ligo o3a data},
  \href{https://doi.org/10.1103/physrevlett.133.111401}{\emph{Physical Review
  Letters} {\bfseries 133} (Sept., 2024) }.

\bibitem{Andres_Carcasona:2024wqk}
M.~Andr\'es-Carcasona, A.~J. Iovino, V.~Vaskonen, H.~Veerm\"ae,
  M.~Mart\'\i{}nez, O.~Pujol\`as et~al., \emph{{Constraints on primordial black
  holes from LIGO-Virgo-KAGRA O3 events}},
  \href{https://doi.org/10.1103/PhysRevD.110.023040}{\emph{Phys. Rev. D}
  {\bfseries 110} (2024) 023040},
  [\href{https://arxiv.org/abs/2405.05732}{{\ttfamily 2405.05732}}].

\bibitem{Capela_2013}
F.~Capela, M.~Pshirkov and P.~Tinyakov, \emph{Constraints on primordial black
  holes as dark matter candidates from star formation},
  \href{https://doi.org/10.1103/physrevd.87.023507}{\emph{Physical Review D}
  {\bfseries 87} (Jan., 2013) }.

\bibitem{Monroy_Rodr_guez_2014}
M.~A. Monroy-Rodríguez and C.~Allen, \emph{The end of the macho era,
  revisited: New limits on macho masses from halo wide binaries},
  \href{https://doi.org/10.1088/0004-637x/790/2/159}{\emph{The Astrophysical
  Journal} {\bfseries 790} (July, 2014) 159}.

\bibitem{Graham_2015}
P.~W. Graham, S.~Rajendran and J.~Varela, \emph{Dark matter triggers of
  supernovae}, \href{https://doi.org/10.1103/physrevd.92.063007}{\emph{Physical
  Review D} {\bfseries 92} (Sept., 2015) }.

\bibitem{Bullock_2017}
J.~S. Bullock and M.~Boylan-Kolchin, \emph{Small-scale challenges to the
  $\lambda$cdm paradigm},
  \href{https://doi.org/10.1146/annurev-astro-091916-055313}{\emph{Annual
  Review of Astronomy and Astrophysics} {\bfseries 55} (Aug., 2017) 343–387}.

\bibitem{Bean_2002}
R.~Bean and J.~Magueijo, \emph{Could supermassive black holes be quintessential
  primordial black holes?},
  \href{https://doi.org/10.1103/physrevd.66.063505}{\emph{Physical Review D}
  {\bfseries 66} (Sept., 2002) }.

\bibitem{Yuan:2023bvh}
G.-W. Yuan, L.~Lei, Y.-Z. Wang, B.~Wang, Y.-Y. Wang, C.~Chen et~al.,
  \emph{{Rapidly growing primordial black holes as seeds of the massive
  high-redshift JWST Galaxies}},
  \href{https://doi.org/10.1007/s11433-024-2433-3}{\emph{Sci. China Phys. Mech.
  Astron.} {\bfseries 67} (2024) 109512},
  [\href{https://arxiv.org/abs/2303.09391}{{\ttfamily 2303.09391}}].

\bibitem{Bird:2016dcv}
S.~Bird, I.~Cholis, J.~B. Mu\~noz, Y.~Ali-Ha\"\i{}moud, M.~Kamionkowski, E.~D.
  Kovetz et~al., \emph{{Did LIGO detect dark matter?}},
  \href{https://doi.org/10.1103/PhysRevLett.116.201301}{\emph{Phys. Rev. Lett.}
  {\bfseries 116} (2016) 201301},
  [\href{https://arxiv.org/abs/1603.00464}{{\ttfamily 1603.00464}}].

\bibitem{Clesse:2016vqa}
S.~Clesse and J.~Garc\'\i{}a-Bellido, \emph{{The clustering of massive
  Primordial Black Holes as Dark Matter: measuring their mass distribution with
  Advanced LIGO}},
  \href{https://doi.org/10.1016/j.dark.2016.10.002}{\emph{Phys. Dark Univ.}
  {\bfseries 15} (2017) 142--147},
  [\href{https://arxiv.org/abs/1603.05234}{{\ttfamily 1603.05234}}].

\bibitem{Sasaki:2016jop}
M.~Sasaki, T.~Suyama, T.~Tanaka and S.~Yokoyama, \emph{{Primordial Black Hole
  Scenario for the Gravitational-Wave Event GW150914}},
  \href{https://doi.org/10.1103/PhysRevLett.117.061101}{\emph{Phys. Rev. Lett.}
  {\bfseries 117} (2016) 061101},
  [\href{https://arxiv.org/abs/1603.08338}{{\ttfamily 1603.08338}}].

\bibitem{Abbott_2016}
B.~Abbott, R.~Abbott, T.~Abbott, M.~Abernathy, F.~Acernese, K.~Ackley et~al.,
  \emph{Observation of gravitational waves from a binary black hole merger},
  \href{https://doi.org/10.1103/physrevlett.116.061102}{\emph{Physical Review
  Letters} {\bfseries 116} (Feb., 2016) }.

\bibitem{Abbott_2016_2}
B.~Abbott, R.~Abbott, T.~Abbott, M.~Abernathy, F.~Acernese, K.~Ackley et~al.,
  \emph{Gw151226: Observation of gravitational waves from a 22-solar-mass
  binary black hole coalescence},
  \href{https://doi.org/10.1103/physrevlett.116.241103}{\emph{Physical Review
  Letters} {\bfseries 116} (June, 2016) }.

\bibitem{Abbott_2016_3}
B.~P. Abbott, R.~Abbott, T.~D. Abbott, M.~R. Abernathy, F.~Acernese, K.~Ackley
  et~al., \emph{The rate of binary black hole mergers inferred from advanced
  ligo observations surrounding gw150914},
  \href{https://doi.org/10.3847/2041-8205/833/1/l1}{\emph{The Astrophysical
  Journal Letters} {\bfseries 833} (Nov., 2016) L1}.

\bibitem{Garc_a_Bellido_2017}
J.~García-Bellido, \emph{Massive primordial black holes as dark matter and
  their detection with gravitational waves},
  \href{https://doi.org/10.1088/1742-6596/840/1/012032}{\emph{Journal of
  Physics: Conference Series} {\bfseries 840} (May, 2017) 012032}.

\bibitem{Franciolini_2022}
G.~Franciolini, V.~Baibhav, V.~De~Luca, K.~K. Ng, K.~W. Wong, E.~Berti et~al.,
  \emph{Searching for a subpopulation of primordial black holes in ligo-virgo
  gravitational-wave data},
  \href{https://doi.org/10.1103/physrevd.105.083526}{\emph{Physical Review D}
  {\bfseries 105} (Apr., 2022) }.

\bibitem{Bhagwat_2021}
S.~Bhagwat, V.~D. Luca, G.~Franciolini, P.~Pani and A.~Riotto, \emph{The
  importance of priors on ligo-virgo parameter estimation: the case of
  primordial black holes},
  \href{https://doi.org/10.1088/1475-7516/2021/01/037}{\emph{Journal of
  Cosmology and Astroparticle Physics} {\bfseries 2021} (Jan., 2021)
  037–037}.

\bibitem{Chen_2018}
Z.-C. Chen and Q.-G. Huang, \emph{Merger rate distribution of primordial black
  hole binaries}, \href{https://doi.org/10.3847/1538-4357/aad6e2}{\emph{The
  Astrophysical Journal} {\bfseries 864} (Aug., 2018) 61}.

\bibitem{Chen_2022}
Z.-C. Chen, C.~Yuan and Q.-G. Huang, \emph{Confronting the primordial black
  hole scenario with the gravitational-wave events detected by ligo-virgo},
  \href{https://doi.org/10.1016/j.physletb.2022.137040}{\emph{Physics Letters
  B} {\bfseries 829} (June, 2022) 137040}.

\bibitem{De_Luca_2021}
V.~De~Luca, V.~Desjacques, G.~Franciolini, P.~Pani and A.~Riotto,
  \emph{Gw190521 mass gap event and the primordial black hole scenario},
  \href{https://doi.org/10.1103/physrevlett.126.051101}{\emph{Physical Review
  Letters} {\bfseries 126} (Feb., 2021) }.

\bibitem{De_Luca_2020}
V.~De~Luca, G.~Franciolini, P.~Pani and A.~Riotto, \emph{Constraints on
  primordial black holes: The importance of accretion},
  \href{https://doi.org/10.1103/physrevd.102.043505}{\emph{Physical Review D}
  {\bfseries 102} (Aug., 2020) }.

\bibitem{Slatyer_2013}
T.~R. Slatyer, \emph{Energy injection and absorption in the cosmic dark ages},
  \href{https://doi.org/10.1103/physrevd.87.123513}{\emph{Physical Review D}
  {\bfseries 87} (June, 2013) }.

\bibitem{Padmanabhan:2005es}
N.~Padmanabhan and D.~P. Finkbeiner, \emph{{Detecting dark matter annihilation
  with CMB polarization: Signatures and experimental prospects}},
  \href{https://doi.org/10.1103/PhysRevD.72.023508}{\emph{Phys. Rev. D}
  {\bfseries 72} (2005) 023508},
  [\href{https://arxiv.org/abs/astro-ph/0503486}{{\ttfamily
  astro-ph/0503486}}].

\bibitem{1952AJ.....57R..31W}
S.~A. {Wouthuysen}, \emph{{On the excitation mechanism of the 21-cm
  (radio-frequency) interstellar hydrogen emission line.}},
  \href{https://doi.org/10.1086/106661}{\emph{\aj} {\bfseries 57} (Jan., 1952)
  31--32}.

\bibitem{4065250}
G.~B. Field, \emph{Excitation of the hydrogen 21-cm line},
  \href{https://doi.org/10.1109/JRPROC.1958.286741}{\emph{Proceedings of the
  IRE} {\bfseries 46} (1958) 240--250}.

\bibitem{2020}
N.~Aghanim, Y.~Akrami, M.~Ashdown, J.~Aumont, C.~Baccigalupi, M.~Ballardini
  et~al., \emph{Planck2018 results: Vi. cosmological parameters},
  \href{https://doi.org/10.1051/0004-6361/201833910}{\emph{Astronomy \&
  Astrophysics} {\bfseries 641} (Sept., 2020) A6}.

\bibitem{10.1093/mnras/104.5.273}
H.~Bondi and F.~Hoyle, \emph{{On the Mechanism of Accretion by Stars}},
  \href{https://doi.org/10.1093/mnras/104.5.273}{\emph{Monthly Notices of the
  Royal Astronomical Society} {\bfseries 104} (10, 1944) 273--282}.

\bibitem{Yang_2021}
Y.~Yang, \emph{Constraints on accreting primordial black holes with the global
  21-cm signal},
  \href{https://doi.org/10.1103/physrevd.104.063528}{\emph{Physical Review D}
  {\bfseries 104} (Sept., 2021) }.

\bibitem{Ricotti_2007}
M.~Ricotti, \emph{Bondi accretion in the early universe}, {\emph{The
  Astrophysical Journal} {\bfseries 662} (jun, 2007) 53}.

\bibitem{Gilfanov_2009}
M.~Gilfanov, \emph{{X-ray emission from black-hole binaries}},
  \href{https://doi.org/10.1007/978-3-540-76937-8_2}{\emph{Lect. Notes Phys.}
  {\bfseries 794} (2010) 17--51},
  [\href{https://arxiv.org/abs/0909.2567}{{\ttfamily 0909.2567}}].

\bibitem{Zhang:2023rnp}
Z.~Zhang, B.~Yue, Y.~Xu, Y.-Z. Ma, X.~Chen and M.~Liu, \emph{{Cosmic radio
  background from primordial black holes at cosmic dawn}},
  \href{https://doi.org/10.1103/PhysRevD.107.083013}{\emph{Phys. Rev. D}
  {\bfseries 107} (2023) 083013},
  [\href{https://arxiv.org/abs/2303.06616}{{\ttfamily 2303.06616}}].

\bibitem{Narayan:2021qfw}
R.~Narayan, A.~Chael, K.~Chatterjee, A.~Ricarte and B.~Curd, \emph{{Jets in
  magnetically arrested hot accretion flows: geometry, power, and black hole
  spin-down}}, \href{https://doi.org/10.1093/mnras/stac285}{\emph{Mon. Not.
  Roy. Astron. Soc.} {\bfseries 511} (2022) 3795--3813},
  [\href{https://arxiv.org/abs/2108.12380}{{\ttfamily 2108.12380}}].

\bibitem{Yuan:2014gma}
F.~Yuan and R.~Narayan, \emph{{Hot Accretion Flows Around Black Holes}},
  \href{https://doi.org/10.1146/annurev-astro-082812-141003}{\emph{Ann. Rev.
  Astron. Astrophys.} {\bfseries 52} (2014) 529--588},
  [\href{https://arxiv.org/abs/1401.0586}{{\ttfamily 1401.0586}}].

\bibitem{Hektor:2018qqw}
A.~Hektor, G.~H\"utsi, L.~Marzola, M.~Raidal, V.~Vaskonen and H.~Veerm\"ae,
  \emph{{Constraining Primordial Black Holes with the EDGES 21-cm Absorption
  Signal}}, \href{https://doi.org/10.1103/PhysRevD.98.023503}{\emph{Phys. Rev.
  D} {\bfseries 98} (2018) 023503},
  [\href{https://arxiv.org/abs/1803.09697}{{\ttfamily 1803.09697}}].

\bibitem{Narayan:2008bv}
R.~Narayan and J.~E. McClintock, \emph{{Advection-Dominated Accretion and the
  Black Hole Event Horizon}},
  \href{https://doi.org/10.1016/j.newar.2008.03.002}{\emph{New Astron. Rev.}
  {\bfseries 51} (2008) 733--751},
  [\href{https://arxiv.org/abs/0803.0322}{{\ttfamily 0803.0322}}].

\bibitem{Merloni:2003aq}
A.~Merloni, S.~Heinz and T.~Di~Matteo, \emph{{A Fundamental plane of black hole
  activity}},
  \href{https://doi.org/10.1046/j.1365-2966.2003.07017.x}{\emph{Mon. Not. Roy.
  Astron. Soc.} {\bfseries 345} (2003) 1057},
  [\href{https://arxiv.org/abs/astro-ph/0305261}{{\ttfamily
  astro-ph/0305261}}].

\bibitem{Lusso:2012yv}
E.~Lusso et~al., \emph{{Bolometric luminosities and Eddington ratios of X-ray
  selected Active Galactic Nuclei in the XMM-COSMOS survey}},
  \href{https://doi.org/10.1111/j.1365-2966.2012.21513.x}{\emph{Mon. Not. Roy.
  Astron. Soc.} {\bfseries 425} (2012) 623},
  [\href{https://arxiv.org/abs/1206.2642}{{\ttfamily 1206.2642}}].

\bibitem{PhysRevD.95.023010}
T.~R. Slatyer and C.-L. Wu, \emph{General constraints on dark matter decay from
  the cosmic microwave background},
  \href{https://doi.org/10.1103/PhysRevD.95.023010}{\emph{Phys. Rev. D}
  {\bfseries 95} (Jan, 2017) 023010}.

\bibitem{PhysRevD.94.063507}
H.~Liu, T.~R. Slatyer and J.~Zavala, \emph{Contributions to cosmic reionization
  from dark matter annihilation and decay},
  \href{https://doi.org/10.1103/PhysRevD.94.063507}{\emph{Phys. Rev. D}
  {\bfseries 94} (Sep, 2016) 063507}.

\bibitem{PhysRevD.102.103005}
J.~Cang, Y.~Gao and Y.-Z. Ma, \emph{Probing dark matter with future cmb
  measurements}, \href{https://doi.org/10.1103/PhysRevD.102.103005}{\emph{Phys.
  Rev. D} {\bfseries 102} (Nov, 2020) 103005}.

\bibitem{Seager:1999bc}
S.~Seager, D.~D. Sasselov and D.~Scott, \emph{{A new calculation of the
  recombination epoch}}, \href{https://doi.org/10.1086/312250}{\emph{Astrophys.
  J. Lett.} {\bfseries 523} (1999) L1--L5},
  [\href{https://arxiv.org/abs/astro-ph/9909275}{{\ttfamily
  astro-ph/9909275}}].

\bibitem{Ali-Haimoud:2010hou}
Y.~Ali-Haimoud and C.~M. Hirata, \emph{{HyRec: A fast and highly accurate
  primordial hydrogen and helium recombination code}},
  \href{https://doi.org/10.1103/PhysRevD.83.043513}{\emph{Phys. Rev. D}
  {\bfseries 83} (2011) 043513},
  [\href{https://arxiv.org/abs/1011.3758}{{\ttfamily 1011.3758}}].

\bibitem{Slatyer_2016}
T.~R. Slatyer, \emph{Indirect dark matter signatures in the cosmic dark ages.
  ii. ionization, heating, and photon production from arbitrary energy
  injections}, \href{https://doi.org/10.1103/physrevd.93.023521}{\emph{Physical
  Review D} {\bfseries 93} (Jan., 2016) }.

\bibitem{shull1985x}
J.~M. Shull and M.~E. Van~Steenberg, \emph{X-ray secondary heating and
  ionization in quasar emission-line clouds}, {\emph{Astrophysical Journal,
  Part 1 (ISSN 0004-637X), vol. 298, Nov. 1, 1985, p. 268-274.} {\bfseries 298}
  (1985) 268--274}.

\bibitem{Galli:2013dna}
S.~Galli, T.~R. Slatyer, M.~Valdes and F.~Iocco, \emph{{Systematic
  Uncertainties In Constraining Dark Matter Annihilation From The Cosmic
  Microwave Background}},
  \href{https://doi.org/10.1103/PhysRevD.88.063502}{\emph{Phys. Rev. D}
  {\bfseries 88} (2013) 063502},
  [\href{https://arxiv.org/abs/1306.0563}{{\ttfamily 1306.0563}}].

\bibitem{Pritchard:2011xb}
J.~R. Pritchard and A.~Loeb, \emph{{21-cm cosmology}},
  \href{https://doi.org/10.1088/0034-4885/75/8/086901}{\emph{Rept. Prog. Phys.}
  {\bfseries 75} (2012) 086901},
  [\href{https://arxiv.org/abs/1109.6012}{{\ttfamily 1109.6012}}].

\bibitem{Pritchard:2006sq}
J.~R. Pritchard and S.~R. Furlanetto, \emph{{21 cm fluctuations from
  inhomogeneous X-ray heating before reionization}},
  \href{https://doi.org/10.1111/j.1365-2966.2007.11519.x}{\emph{Mon. Not. Roy.
  Astron. Soc.} {\bfseries 376} (2007) 1680--1694},
  [\href{https://arxiv.org/abs/astro-ph/0607234}{{\ttfamily
  astro-ph/0607234}}].

\bibitem{Mesinger:2010ne}
A.~Mesinger, S.~Furlanetto and R.~Cen, \emph{{21cmFAST: A Fast, Semi-Numerical
  Simulation of the High-Redshift 21-cm Signal}},
  \href{https://doi.org/10.1111/j.1365-2966.2010.17731.x}{\emph{Mon. Not. Roy.
  Astron. Soc.} {\bfseries 411} (2011) 955},
  [\href{https://arxiv.org/abs/1003.3878}{{\ttfamily 1003.3878}}].

\bibitem{Cang:2024gtp}
J.~Cang, A.~Mesinger, S.~G. Murray, D.~Breitman, Y.~Qin and R.~Trotta,
  \emph{{The EDGES measurement disfavors an excess radio background during the
  cosmic dawn}},  \href{https://arxiv.org/abs/2411.08134}{{\ttfamily
  2411.08134}}.

\bibitem{Tiwari:2024ivw}
H.~Tiwari, N.~Thyagarajan, C.~M. Trott, B.~McKinley and B.~McKinley,
  \emph{{21-cm Epoch of reionisation power spectrum with closure phase using
  the Murchison Widefield Array}},
  \href{https://doi.org/10.1017/pasa.2024.81}{\emph{Publ. Astron. Soc.
  Austral.} {\bfseries 41} (2024) e069},
  [\href{https://arxiv.org/abs/2409.02906}{{\ttfamily 2409.02906}}].

\bibitem{Carucci:2024qpm}
I.~P. Carucci et~al., \emph{{Hydrogen intensity mapping with MeerKAT:
  Preserving cosmological signal by optimising contaminant separation}},
  \href{https://arxiv.org/abs/2412.06750}{{\ttfamily 2412.06750}}.

\bibitem{SKA:2018ckk}
{\scshape SKA} collaboration, D.~J. Bacon et~al., \emph{{Cosmology with Phase 1
  of the Square Kilometre Array: Red Book 2018: Technical specifications and
  performance forecasts}},
  \href{https://doi.org/10.1017/pasa.2019.51}{\emph{Publ. Astron. Soc.
  Austral.} {\bfseries 37} (2020) e007},
  [\href{https://arxiv.org/abs/1811.02743}{{\ttfamily 1811.02743}}].

\bibitem{Carr:2020xqk}
B.~Carr and F.~Kuhnel, \emph{{Primordial Black Holes as Dark Matter: Recent
  Developments}},
  \href{https://doi.org/10.1146/annurev-nucl-050520-125911}{\emph{Ann. Rev.
  Nucl. Part. Sci.} {\bfseries 70} (2020) 355--394},
  [\href{https://arxiv.org/abs/2006.02838}{{\ttfamily 2006.02838}}].

\bibitem{inoue2017new}
Y.~Inoue and A.~Kusenko, \emph{New x-ray bound on density of primordial black
  holes}, {\emph{Journal of Cosmology and Astroparticle Physics} {\bfseries
  2017} (2017) 034}.

\bibitem{niikura2019microlensing}
H.~Niikura, M.~Takada, N.~Yasuda, R.~H. Lupton, T.~Sumi, S.~More et~al.,
  \emph{Microlensing constraints on primordial black holes with subaru/hsc
  andromeda observations}, {\emph{Nature Astronomy} {\bfseries 3} (2019)
  524--534}.

\bibitem{Finkbeiner:2011dx}
D.~P. Finkbeiner, S.~Galli, T.~Lin and T.~R. Slatyer, \emph{{Searching for Dark
  Matter in the CMB: A Compact Parameterization of Energy Injection from New
  Physics}}, \href{https://doi.org/10.1103/PhysRevD.85.043522}{\emph{Phys. Rev.
  D} {\bfseries 85} (2012) 043522},
  [\href{https://arxiv.org/abs/1109.6322}{{\ttfamily 1109.6322}}].

\bibitem{Qin:2024cna}
Y.~Qin et~al., \emph{{Percent-level timing of reionization: self-consistent,
  implicit-likelihood inference from XQR-30+ Ly$\alpha$ forest data}},
  \href{https://arxiv.org/abs/2412.00799}{{\ttfamily 2412.00799}}.

\bibitem{Dai:2018nce}
W.-M. Dai, Y.-Z. Ma, Z.-K. Guo and R.-G. Cai, \emph{{Constraining the
  reionization history with CMB and spectroscopic observations}},
  \href{https://doi.org/10.1103/PhysRevD.99.043524}{\emph{Phys. Rev. D}
  {\bfseries 99} (2019) 043524},
  [\href{https://arxiv.org/abs/1805.02236}{{\ttfamily 1805.02236}}].

\end{thebibliography}\endgroup
\bibliographystyle{JHEP} 

\end{document}